\documentclass[prd,preprintnumbers,nofootinbib]{revtex4} 
\usepackage{graphicx} 
\usepackage{amsmath}
\usepackage{amsfonts,amsbsy}
\usepackage{amssymb}
\usepackage{appendix}
\usepackage{slashed}

\def\be{\begin{equation}}
\def\ee{\end{equation}}
\def\bea{\begin{eqnarray}}
\def\eea{\end{eqnarray}}

\def\eq#1{{Eq.\,(\ref{#1})}}

\def\k{{\mathbf k}}

\def\p{{\mathbf p}}
\def\r{{\mathbf r}}
\def\b{{\mathbf b}}
\def\q{{\mathbf q}}
\def\k{{\mathbf k}}

\def\x{{\mathbf x}}
\def\y{{\mathbf y}}
\def\R{{\mathbf R}}

\begin{document}

\title{\bf DPS in  CGC: Double Quark Production and Effects of Quantum Statistics}


\author{Alex Kovner$^{1,2,3}$ and Amir H. Rezaeian$^{2,3}$}

\affiliation{
$^1$ Dept. of Physics, University of Connecticut, Storrs, CT 06269, USA\\
$^2$ Departamento de F\'\i sica, Universidad T\'ecnica
Federico Santa Mar\'\i a, Avda. Espa\~na 1680,
Casilla 110-V, Valparaiso, Chile\\
$^3$  Centro Cient\'\i fico Tecnol\'ogico de Valpara\'\i so (CCTVal), Universidad T\'ecnica
Federico Santa Mar\'\i a, Casilla 110-V, Valpara\'\i so, Chile
}

\begin{abstract}
We consider forward inclusive production of two quarks in the  high energy  p-A collisions in the CGC formalism. We demonstrate that the production cross-section is determined by the convolution of the  proton  generalized double transverse momentum-dependent distribution (2GTMD) functions with two independent eikonal scattering amplitudes: the  product of two dipoles and a quadrupole.   We explicitly demonstrate that the quadrupole amplitude term accounts for all the (initial and final state) effects of quantum statistics for identical fermions, and the correlations due to these effects. We also demonstrate that the effects due to quantum statistics (entirely encoded in the quadrupole) are parametrically leading contributions to the correlated particle production  at large $N_c$. For non-identical quarks the quadrupole term also leads to correlated production which (barring accidental cancellations) has characteristics similar to the HBT effect.
 
\end{abstract}

\maketitle

\section{Introduction}
In recent years there has been a growing interest in correlated particle production in high energy collisions, triggered mainly by the observation of ridge correlations  in p-p and p-Pb collisions at LHC \cite{exp-r1,exp-r2,exp-r3,exp-r4,exp-r5}. Currently these correlations are explained either by the strong collective effects due to final state interactions \cite{hydro}, or due to ``quasi collectivity'' - correlations inherited from the nontrivial correlated structure of the initial state \cite{ridge0, ridge1,ridge2,ridge3,ridge1-raju,ridge33,ridge4,ridge5}.

 One of the spin-offs of the activity aimed at understanding the ridge has been a closer study of the correlations due to quantum statistics of observed particles. In particular in the framework of CGC it has been shown that the correlations appearing in the so-called ``glasma graph'' calculation are mostly due to Bose enhancement of gluons in the high energy proton wave function \cite{bosee}, with another component due to the Hanbury Brown-Twiss (HBT)  effect in the emission of identical gluons \cite{hbthadrons}.  Subsequently Pauli blocking effect relevant for quark production at mid rapidity  has been studied in the CGC fomralizm \cite{paulib}, and interesting correlated structure has been observed also for gluon-two quark correlations \cite{2qg}.

More recently we have studied the HBT effect for photon emission in p-A collisions at forward rapidity \cite{kr1,kr}. We observed that this effect can be used for extracting the nonperturbative proton size scale from the generalized two particle transverse momentum dependent distributions (2GTMD)  \cite{kr1}. In the present paper we continue our study of effects of quantum statistics in the forward particle production. We consider the simple case of production of two quarks at forward rapidity within the CGC based formalism. Although this is a very simple and seemingly even trivial exercise, our results turn out to be quite illuminating and shed some new light on the production of identical particles, and the origin of correlations in CGC-like calculations.

In the present paper we use  the extension of the hybrid formalism to include the double-parton-scattering (DPS) in the Color-Glass-Condensate (CGC) approach \cite{kr1}.  
We will be working within a variant of the "hybrid" approximation \cite{dj-rhic} which is appropriate for forward particle production. In the hybrid  CGC approach, we assume that the small-x gluon modes of the nucleus have a large occupation number so that the target nucleus can be described in terms of a classical color field.  This should be a good approximation for large enough nucleus at high-energy. This color field emerges from the classical Yang-Mills equation with a source term provided by faster partons. The renormalization group equations which govern the separation between the soft and hard models are then given by the non-linear Jalilian-Marian, Iancu, McLerran, Weigert, Leonidov, Kovner (JIMWLK) \cite{jimwlk} and Balitsky-Kovchegov (BK) \cite{bk} evolution equations.  We further assume that the projectile proton is in the dilute regime and can be described in ordinary perturbative approach using parton picture like assumptions. This somewhat restricts the validity range of our approximation, and we  comment on the pertinent points in the text and in the Appendix.

Our results can be summarized as follows. The double inclusive quark production cross-section can be quite generally expressed in terms of the two particle generalized transverse momentum dependent distribution (2GTMD) convoluted with a function of eikonal scattering factors ubiquitous in the CGC based calculations. In the context of the collinear factorization,  objects similar to 2GTMD, the so-called generalized double parton distribution (2GPD) appears in studies of the DPS \cite{marke}, see also Refs.\,\cite{2gpd-1,2gpd-2,2gpd-3,2gpd-4,2gpd-5}. As for the functions of eikonal factors, generally speaking there are two such functions that appear in the calculation - the product of two dipoles and a quadrupole. We calculate the double quark production cross section for nonidentical ($ud$) and identical ($uu$) quarks.  We demonstrate that for identical particles  the quadrupole term accounts for all the quantum statistics effects. We also show explicitly that the Pauli blocking and the Hanbury-Brown, Twiss (HBT) correlations originate from the part of the phase space where the quadrupole amplitude can be reduced to product of two dipoles without recourse to weak target  field limit.
Surprisingly, we find that even for nonidentical quarks the quadrupole term leads to correlated production which at least superficially has characteristics similar to HBT suppression.  Quantitative study of this effect is beyond this paper, since it involves understanding of (nonperturbative)  off diagonal analog of quark transverse dependent momentum distribution (see below).

  To simplify our general expressions, we  use a common and physically motivated assumption that the proton 2GTMD is dominated by the nucleon intermediate states, and can therefore be factorized into product of two single quark GTMD's. We further analyze the quadrupole  term and confirm explicitly that it yields both quantum statistics effects mentioned above, the Pauli blocking and the HBT correlations as contributions coming from different parts of the quadrupole phase space which is integrated over in the calculation of the cross-section. Furthermore we show that the correlation effects contained in the two dipole term are {\bf suppressed} in the large $N_c$ limit relatively to the correlations due to quantum statistics by a factor of $1/N_c$. Thus any attempt to ascribe experimentally observed correlations to the structure of initial state {\bf must} include the effects of quantum statistics, and therefore the quadrupole term in the production cross-section. 

We note that some recent  calculations of particle correlations \cite{dipoles} neglect the quadrupole contribution altogether, and thus miss the leading $N_c$ contribution to this observable. The effects of quantum statistics in multiple particle production compete with the effects of the type studied in \cite{dipoles} as explained in \cite{vladi} for the case of gluons. The balance between the two types of effects has to be studied carefully in a quantitative manner before definitive conclusions about the strength of correlations can be reached.

This paper is organized as follows: In Sec. II, we provide a concise description of theoretical framework for calculating the DPS contribution in the CGC approach, and derive the general expressions for the two quark inclusive cross-section.  In Sec. III we discuss the approximation to 2GTMDs assuming the dominance of the nucleon intermediate state, and simplify our general formula in this approximation. 
In Sec. IV we show explicitly how the Pauli blocking and the HBT effect arise from the quadrupole term in the cross-section for identical quarks, and comment on the magnitude of correlations arising from the dipole term. Finally Sec. V contains a short discussion on the relation of our results to those of the ``glasma graph'' calculations prevalent in the literature.

\begin{figure}[t]                                       
                                  \includegraphics[width=9 cm] {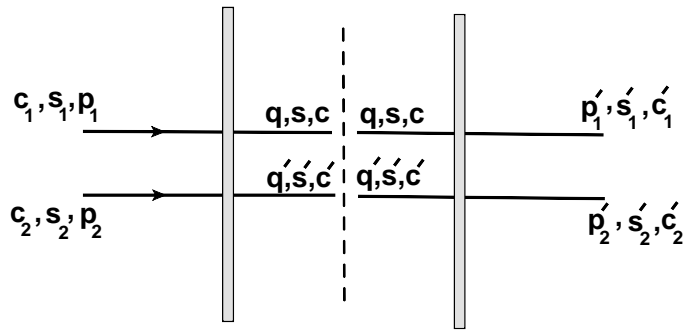}             
\caption{The diagram contributing to two quark production  in the background of the CGC field. The diagrams on the left and right side of the dashed line correspond to the amplitude and the complex conjugate amplitude.  The shaded box (the CGC shock waive) denotes the interaction of a quark to all orders with the background field via multiple gluon exchanges.  The color ($c_i, c'_i$), spin ($s_i, s'_i$) and momenta ($p_i, p'_i$) of the $i$-th quark in the amplitude and the complex conjugate amplitude  are also shown.  }
\label{f2}
\end{figure}
\section{Semi-inclusive diquark production in proton-nucleus collisions}
The cross-section for production of two quarks with momentum $q$ and $q^\prime$ in the proton-nucleus (p-A) scatterings  can be written in the following general form, 
\be d\, \sigma^{p+A\to qq+X} =  \frac{d^3 q}{(2\pi)^3\, 2
  q^-} \frac{d^3 q^\prime}{(2\pi)^3\, 2
  q^{\prime -}} 
  \ \langle| \langle \text{jet}(q), \text{jet}(q^\prime)| \text{Proton} \rangle|^2 \rangle_{\text{color sources}}, 
\label{m1}
\ee
where $|\text{Proton}\rangle$ is the wave function of the energetic  proton with vanishing transverse momentum, and the averaging should be performed over the distribution of the color charges in the target. The proton wave function is normalized to unity, $\langle \text{Proton}|\text{Proton}\rangle=1$.

We start by writing the wave function of the proton in the following most general form \cite{kr1}, 
\begin{equation}
|\text{Proton} \rangle=\sum_X\sum_{c_1,s_1}\sum_{c_2,s_2}\int\int\frac{d^3p_1}{(2\pi)^3}\frac{d^3p_2}{(2\pi)^3}\tilde A(p_1,c_1,s_1;p_1,c_2,s_2;X)|p_1,c_1,s_1;p_2,c_2,s_2;X\rangle, 
\end{equation}
where $(p_i,c_i,s_i)$ labels the momentum, color and spin of the $i$-th quark, and $X$ labels the configuration of all the spectator partons in the proton. The spectator particles $X$ also take part in the scattering process and produce particles in the final state. However their contribution to the double inclusive cross-section in the regime of validity of our calculation cancels due to unitarity. This is explicit in our derivation and is discussed in more detail in the Appendix.
The two quarks can be either distinguishable (have different flavors), or indistinguishable (same flavor). We will consider both cases in the following. 

The S-matrix element of the proton scattering into the state with two quarks and an arbitrary configuration of spectator particles can be written as
\begin{eqnarray}
\langle q,c,s;q',c',s';X'|\text{Proton} \rangle&=&\sum_X\sum_{c_1,s_1}\sum_{c_2,s_2}\int\int\frac{d^3p_1}{(2\pi)^3}\frac{d^3p_2}{(2\pi)^3} \nonumber\\
&\times&\tilde A(p_1,c_1,s_1;p_2,c_2,s_2;X)\langle q,c,s;q',c',s';X'|p_1,c_1,s_1;p_2,c_2,s_2;X\rangle.  \
\end{eqnarray}
In the above $\langle A|B\rangle$ denotes the S-matrix element of the initial state $B$ scattering  into the final state $A$. In the spirit of the parton model we assume that partons scatter independently. For two distinct quarks this translates into
\begin{equation}
\langle q,c,s;q',c',s';X'|p_1,c_1,s_1;p_2,c_2,s_2;X\rangle=\langle q,c,s|p_1,c_1,s_1\rangle \langle q',c',s'|p_2,c_2,s_2\rangle\langle X'|X\rangle, 
\end{equation}
while for identical quarks we have
\begin{eqnarray}
\langle q,c,s;q',c',s';X'|p_1,c_1,s_1;p_2,c_2,s_2;X\rangle&=&\Big[\langle q,c,s|p_1,c_1,s_1\rangle \langle q',c',s'|p_2,c_2,s_2\rangle -\langle q,c,s|p_2,c_2,s_2\rangle \langle q',c',s'|p_1,c_1,s_1\rangle  \Big] \nonumber\\
&\times& \langle X'|X\rangle. 
\end{eqnarray}
We will now define for the identical quark case the symmetrized amplitude
\begin{equation}
 A(p_1,c_1,s_1;p_2,c_2,s_2;X)\equiv \tilde  A(p_1,c_1,s_1;p_1,c_2,s_2;X)-\tilde  A(p_2,c_2,s_2;p_1,c_1,s_1;X), 
\end{equation}
while for distinct quarks
\begin{equation}
 A(p_1,c_1,s_1;p_2,c_2,s_2;X)\equiv \tilde  A(p_1,c_1,s_1;p_1,c_2,s_2;X). 
 \end{equation}
With the above definitions we can write down the expression for the double  inclusive cross-section  as
\begin{eqnarray}\label{cross}
\mathcal{I}=\sum_{X'} |\langle q,c,s;q',c',s';X'|\text{Proton}\rangle|^2&=&\sum_{c_1,c_2,c'_1,c'_2,c,c'}\sum_{s_1,s_2,s'_1,s'_2,s,s'}\int_{p_1,p_2,p'_1,p'_2}\sum_X \nonumber\\ &\times& A(p_1,c_1,s_1;p_2,c_2,s_2;X)A^*(p'_1,c'_1,s'_1;p'_2,c'_2,s'_2;X)\\
&\times& \Bigg[\langle q,c,s|p_1,c_1,s_1\rangle \langle q,c,s|p'_1,c'_1,s'_1\rangle^* \Bigg]
\Bigg[\langle q',c',s'|p_2,c_2,s_2\rangle
 \langle q',c',s'|p'_2,c'_2,s'_2\rangle^*\Bigg].\nonumber
\end{eqnarray}
The above expression is only valid under the parton model assumption, namely for the cases that the typical transverse momentum of the quarks in the proton wave function is much smaller than the momentum of the produced particles. If this is not the case, additional terms arise in the expression for the cross-section. The evaluation of these extra terms requires the knowledge of complicated matrix elements, which goes beyond our present ability. This issue is discussed in the Appendix. Throughout this paper we therefore limit ourselves to consideration of large transverse momentum of produced quarks.

For the single quark scattering amplitude we have \cite{gf}, 
\begin{equation}\label{ampl}
\langle q,c,s|p_1,c_1,s_1\rangle=2\pi\delta(p_1^{-}-q^-)\frac{1}{\sqrt{2p^-_1}} \int d^2\x e^{i(\ p_1-\q)\x}[U(\x)]_{cc_1}\bar u_s(q)\gamma^-u_{s_1}(p_1), 
\end{equation}
where $U(\x)$ is the scattering matrix of a quark on the colored glass condensate target, and it is represented as a unitary matrix in fundamental representation of $SU(N_c)$. 
The factor $\frac{1}{\sqrt{2p^-_1}}$ was introduced for convenience in order to avoid extra normalization factor in the cross-section defined in \eq{m1}.  
Throughout the paper we denote transverse coordinates and momenta by boldface letters.  In the following we use the standard relation between spinors, namely $\bar u_s(q)u_{s'}(q)=\slashed{q}_{ss'}$. Using Eq.\,(\ref{ampl}), the cross-section is written as
\begin{eqnarray}
\mathcal{I}&=&(2\pi)^4\delta(p_1^{-}-q^-)\delta(p_1^{\prime -}-q^-)\delta(p_2^--q'^{-})\delta(p_2^{\prime -}-q'^{-})\frac{1}{4qq'}\sum_{c_1,c_2,c'_1,c'_2,c,c'}\sum_{s_1,s_2,s'_1,s'_2}\int_{\p_1,\p_2,\p'_1,\p'_2}\nonumber\\
&&\times\langle[U^\dagger(\x_1)]_{c'_1c}[U(\x_2)]_{cc_1}[U^\dagger(\x_3)]_{c'_2c'}[U(\x_4)]_{c'c_2}\rangle e^{i[(\p'_1-\q)\x_1+(\q-\p_1)\x_2+(\p'_2-\q')\x_3+(\q'-\p_2)\x_4]}\nonumber\\
&&\bar u_{s'_1}(p'_1)\gamma^-\slashed{q}\gamma^- u_{s_1}(p_1)\bar u_{s'_2}(p'_2)\gamma^-\slashed{q}'\gamma^-u_{s_2}(p_2)\sum_X A(p_1,c_1,s_1;p_2,c_2,s_2;X)A^*(p'_1,c'_1,s'_1;p'_2,c'_2,s'_2;X),  \label{m-2}\
\end{eqnarray}
where for brevity, we used a notation  $\int \frac{d^2\p}{(2\pi)^2}\equiv \int_p$. 

In the high energy limit one can perform the spin algebra in a straightforward way.
First, 
we have
\begin{equation}
\gamma^-\slashed{q}\gamma^-= 2\gamma^-q^-. 
\end{equation}
In the  approximation where the largest component of momentum $p_i^\mu$ is $p_i^-$ (and the same for $p_i'$)
 the spinors do not depend on the transverse momentum, so that for different momenta  they only differ by a normalization factor 
$\frac{1}{\sqrt{p^-_1}}u^{s}(p_1)=\frac{1}{\sqrt{p^{'-}_1}}u^{s}(p'_1)$.  Therefore, at high energies we have
\begin{equation}
\bar u_{s'_1}(p'_1)\gamma^-\slashed{q}\gamma^- u_{s_1}(p_1)\bar u_{s'_2}(p'_2)\gamma^-\slashed{q}'\gamma^-u_{s_2}(p_2)=16 q^-q'^{-}\sqrt{p_1^-p_2^-p'^{-}_1 p^{\prime -}_2}
\delta_{s_1s'_1}\delta_{s_2s'_2}.
\end{equation}
Using the above relation, we can simplify the spin algebra in Eq.\, (\ref{m-2}) and obtain 
\begin{eqnarray}
\mathcal{I}&=&16\frac{(2\pi)^4}{4}\delta(p_1^{-}-q^-)\delta(p_1^{\prime -}-q^-)\delta(p_2^--q'^{-})\delta(p_2^{\prime -}-q'^{-})q^-q'^{-}\sum_{c_1,c_2,c'_1,c'_2,c,c'}\sum_{s_1,s_2}\int_{\p_1,\p_2,\p'_1,\p'_2}\nonumber\\
&&\times\langle[U^\dagger(\x_1)]_{c'_1c}[U(\x_2)]_{cc_1}[U^\dagger(\x_3)]_{c'_2c'}[U(\x_4)]_{c'c_2}\rangle e^{i[(\p'_1-\q)\x_1+(\q-\p_1)\x_2+(\p'_2-\q')\x_3+(\q'-\p_2)\x_4]}\nonumber\\
&&\sum_X A(p_1,c_1,s_1;p_2,c_2,s_2;X)A^*(p'_1,c'_1,s_1;p'_2,c'_2,s_2;X). 
\end{eqnarray}

\subsection{Color algebra}
The averaging over the eikonal scattering matrices has to be performed over the target ensemble. We will use the fact that the target ensemble has to be color invariant.The average of any tensor  in such an ensemble must be proportional to a linear combination  of available invariant tensors. Consequently for any such ensemble we must have
\begin{equation}
\langle[U^\dagger(\x_1)]_{c'_1c}[U(\x_2)]_{cc_1}[U^\dagger(\x_3)]_{c'_2c'}[U(\x_4)]_{c'c_2}\rangle=\mathcal{A} \delta_{c'_1c_1}\delta_{c'_2c_2}+\mathcal{B}\delta_{c'_1c_2}\delta_{c'_2c_1}.
\end{equation}
Multiplying this relation by $\delta_{c'_1c_1}\delta_{c'_2c_2}$ and then by $\delta_{c'_1c_2}\delta_{c'_2c_1}$ we get two simple linear equations for $\mathcal{A}$ and $\mathcal{B}$ which yield:
\begin{eqnarray}
\mathcal{A}&=&\frac{1}{N_c^2-1}\left[N_c^2D(\x_1,\x_2)D(\x_3,\x_4)-Q(\x_1,\x_2,\x_3,\x_4)\right],\nonumber\\
\mathcal{B} &=&\frac{N_c}{N_c^2-1}\left[Q(\x_1,\x_2,\x_3,\x_4)-D(\x_1,\x_2)D(\x_3,\x_4)\right],
\end{eqnarray}
with
\begin{eqnarray}
D(\x_1,\x_2)&\equiv&\frac{1}{N_c}Tr\Big[U^\dagger(\x_1)U(\x_2)\Big],\nonumber\\
Q(\x_1,\x_2,\x_3,\x_4)&\equiv&\frac{1}{N_c}Tr \Big[U^\dagger(\x_1)U(\x_2)U^\dagger(\x_3)U(\x_4)\Big],
\end{eqnarray}
where $D$ and $Q$ are the traces of two (dipole) and four (quadrupole) light-like fundamental Wilson lines
in the background of the color fields of the target respectively.  
It is now convenient to rewrite the expressions in the momentum space.
To this end we write $\x_{1-4}$ in terms of $\r=\x_1-\x_2$, $\r^\prime=\x_3-\x_4$, $\R=\frac{\x_1+\x_2}{2}-\frac{\x_3+\x_4}{2}$ and $\b=\frac{1}{2}(\x_1+\x_2+\x_3+\x_4)$ where $\b$ is the impact parameter and $\r,r^\prime$ are the separation distance vectors between a quark and an anti quark, and $\R$ is the distance between center-of-mass of the two dipoles:
\begin{eqnarray}
\x_1&=&\frac{1}{2}(\R+\b+\r),\nonumber\\
\x_2&=&\frac{1}{2}(\R+\b-\r),\nonumber\\
\x_3&=&\frac{1}{2}(\r^\prime+\b-\R),\nonumber\\
\x_4&=&\frac{1}{2}(\b-\R-\r^\prime).\
\end{eqnarray}
Now performing integrals over $\r,\r^\prime, \b$ and $\R$, and assuming that the target is translationally  invariant we obtain, 
\begin{eqnarray}
\mathcal{I}&=& 16\frac{(2\pi)^4}{4}\delta(p_1^{-}-q^-)\delta(p_1^{\prime -}-q^-)\delta(p_2^--q'^{-})\delta(p_2^{\prime -}-q'^{-})q^-q'^{-} \nonumber\\
&&\frac{N_c^2}{N_c^2-1}\int_{\p_1,\p_2,\Delta}\Big[ \langle D(\p_1-\q+\Delta/2,2\Delta)D(\p_2-\q^\prime-\Delta/2,-2\Delta ) \rangle  \nonumber\\
 &\times&\sum_{c_1,s_1,c_2,s_2}  \{A(p_1,c_1,s_1;p_2,c_2,s_2;X)A^*(p_1+\Delta,c_1,s_1;p_2-\Delta,c_2,s_2;X)\nonumber\\
 &-& \frac{1}{N_c} A(p_1,c_1,s_1;p_2,c_2,s_2;X)A^*(p_1+\Delta,c_2,s_1;p_2-\Delta,c_1,s_2;X)\} \nonumber\\
 &+&\frac{1}{N_c}\langle Q(\p_1-\q+\Delta/2,\p_2-\q^\prime-\Delta/2, \Delta)\rangle  \sum_{c_1,s_1,c_2,s_2}  \{ A(p_1,c_1,s_1;p_2,c_2,s_2;X)A^*(p_1+\Delta,c_2,s_1;p_2-\Delta,c_1,s_2;X)\nonumber\\
 &-&\frac{1}{N_c}A(p_1,c_1,s_1;p_2,c_2,s_2;X)A^*(p_1+\Delta,c_1,s_1;p_2-\Delta,c_2,s_2;X)\}\Big], \label{m-3}\
 \end{eqnarray}
 where $\Delta=\p^\prime_1-\p_1=\p_2-\p^\prime_2$ is the difference of the momenta of two quarks from the wave function of the colliding hadron in the amplitude and the conjugate amplitude.  
 In the above the Fourier transforms of the dipole and quadrupole amplitudes are defined as
 \begin{eqnarray}\label{ft}
 D(\p,\k)&\equiv&\int d^2\x_1d^2\x_2e^{i\p(\x_1-\x_2)+i\k\frac{\x_1+\x_2}{2}}D(\x_1,\x_2),\\
 Q(\p,\k,\q)&\equiv&\int d^2\r d^2\r' d^2\R e^{i\p(\x_1-\x_2)+i\k(\x_3-\x_4)+i\q\frac{\x_1+\x_2-\x_3-\x_4}{2}}Q(\x_1,\x_2,\x_3,\x_4).\nonumber
 \end{eqnarray} 
 The physical meaning of different terms in Eq.\,(\ref{m-3}) is further clarified by rewriting  \eq{m-3} in terms of the direct and exchange contributions defined by\footnote{The 2GTMD's defined in Eq.(\ref{d-e}) are clearly  invariant under the global (${\bf x}$-independent) gauge transformation. A more careful analysis should be performed to determine the projectile Wilson line factors that ensure their local gauge invariance. We leave the determination of these factors for future work.}: 
  \begin{eqnarray}
&&T^D_{qq'}(p_1,p_2,\Delta)\equiv \sum_{s_1,s_2,c_1,c_2}\sum_X A(\p_1,x_1,c_1,s_1;\p_2,x_2,c_2,s_2;X) A^*(\p_1+\Delta,x_1,c_1,s_1;\p_2-\Delta,x_2,c_2,s_2;X),\nonumber\\
&=& \sum_{s_1,s_2,c_1,c_2}
 \langle \text{Proton}| \psi^{\dagger}_q(\p_1,x_1,c_1,s_1)  \psi^{\dagger}_{q'}(\p_2,x_2,c_2,s_2) \psi_{q'}(\p_2-\Delta,x_2,c_2,s_2) \psi_q(\p_1+\Delta,x_1,c_1,s_1) |\text{Proton} \rangle, \nonumber\\
 &&T^E_{qq'}(p_1,p_2,\Delta)\equiv \sum_{s_1,s_2,c_1,c_2} \sum_X A(\p_1,x_1,c_1,s_1;\p_2,x_2,c_2,s_2;X) A^*(\p_1+\Delta,x_1,c_2,s_1;\p_2-\Delta,x_2,c_1,s_2;X),\nonumber\\
 &=&  \sum_{s_1,s_2,c_1,c_2}\langle \text{Proton}| \psi^{\dagger}_q(\p_1,x_1,c_1,s_1)  \psi^{\dagger}_{q'}(\p_2,x_2,c_2,s_2) \psi_{q'}(\p_2-\Delta,x_2,c_1,s_2) \psi_q(\p_1+\Delta,x_1,c_2,s_1) |\text{Proton}\rangle, \label{d-e}\
 \end{eqnarray}
 where $\psi_q$ and $\psi^\dagger_q$ are quark annihilation and creation operators with a flavor $q$ (e.g. $q=u,d$). In the following we consider the cases of distinct  quark production $q=u,\ q'=d$ and identical quark production $q=q'=u$. In the above expressions we have explicitly indicated the longitudinal momentum fraction of the two quarks. Using the above definitions, we then simplify the expression in \eq{m-3}  and obtain
 \begin{eqnarray}\label{IDE}
\mathcal{I}&=& 16\frac{(2\pi)^4}{4}\delta(p_1^{-}-q^-)\delta(p_1^{\prime -}-q^-)\delta(p_2^--q'^{-})\delta(p_2^{\prime -}-q'^{-})q^-q'^{-}\nonumber\\
&&\frac{N_c^2}{N_c^2-1}\int_{\p_1,\p_2,\Delta}\Big\{ \langle D(\p_1-\q+\Delta/2,2\Delta)D(\p_2-\q^\prime-\Delta/2,-2\Delta ) \rangle\Big[ T^D_{qq'}(p_1,p_2,\Delta)-\frac{1}{N_c}T^{E}_{qq'}(p_1,p_2,\Delta)\Big] \nonumber\\
 &+&\frac{1}{N_c}\langle Q(\p_1-\q+\Delta/2,\ \p_2-\q^\prime-\Delta/2, \Delta)\rangle  \Big[ T^E_{qq'}(p_1,p_2,\Delta)-\frac{1}{N_c}T^D_{qq'}(p_1,p_2,\Delta)\Big] \Big\}, 
 \end{eqnarray}
 The quantities $T^D_{qq'}$ and $T^E_{qq'}$ are off-diagonal two particle densities, and as such fairly complicated objects.  These are nonperturbative quantities that characterize the proton wave function. While the expression for the two quark production is general, it is not very informative as it stands, since it is parametrized by these two nonperturbative objects.
 Eq.\,(\ref{IDE}) is as far as we can get without making some simplifying assumption about their nature.
  We can however extract usable information from it if we employ some simplifying assumptions  about the two particle density. In particular a simple and physically motivated assumption in the spirit of the parton model  is that the two particle distributions are independent of each other, and that the two particle density is dominated by single proton intermediate state. In the next section we will explore this simple assumption.
 
 \section{Dominance of the nucleon intermediate states}
 In this section, we examine the contributions of the direct $T^D_{qq'}$ and the exchange $T^E_{qq'}$ terms defined in \eq{d-e} 
 in the approximation that single nucleon intermediate states  dominate the two particle density average. 
 
 \subsection{Non-identical quarks}

 Consider first the simpler case when the two quarks are distinguishable, i.e. have different flavor. 
  Let us first consider the direct term. 
 Quite generally using the resolution of identity we can write
 \begin{eqnarray}\label{resolu}
 && \sum_{ab;s_1s_2} \langle \text{Proton}| \psi_u^{\dagger}(\p_1,x_1,a,s_1)  \psi_d^{\dagger}(\p_2,x_2,b,s_2) \psi_d(\p_2-\Delta,x_2,b,s_2) \psi_u(\p_1+\Delta,x_1,a,s_1) |\text{Proton} \rangle, \nonumber\\
 &&= \sum_i \langle \text{Proton}|\sum_{a,s_1} \psi_u^{\dagger}(\p_1,x_1,a,s_1)  \psi_u(\p_1+\Delta,x_1,a,s_1) |i\rangle\langle i|\sum_{b,s_2}\psi_d^{\dagger}(\p_2,x_2,b,s_2) \psi_d(\p_2-\Delta,x_2,b,s_2) |\text{Proton} \rangle, \nonumber\\
  \end{eqnarray}
 where the summation goes over all intermediate hadronic states $| i\rangle$, which in principle include  sates with ``inelastically scattered'' proton. However, as long as the momentum transfer from the density operator is small, the main effect of acting with $\sum_a \psi_u^{\dagger}(\p_1,x_1,a,s_1)  \psi_u(\p_1+\Delta,x_1,a,s_1)$ on the original proton state $|\text{Proton} \rangle$ is not to change its nature significantly, but rather to scatter it elastically and impart to it a recoil momentum equal to the total momentum of the density operator. Recall that within our parton model like approximation the momenta of the quark operators are assumed to be of the typical hadronic scale size rather than the relatively high momentum of the produced particles. Thus it is reasonable to expect that the intermediate proton states may dominate the two particle density. 
 
 Note that we have inserted the resolution of identity between the color singlet operators in eq.(\ref{resolu}). This ensures that the nonvanishing contribution comes only from physical color singlet states.
 
 The dominance of proton intermediate states then means
 \begin{eqnarray}\label{pd}
&&\sum_{ab;s_1s_2} \langle \text{Proton}| \psi_u^{\dagger}(\p_1,x_1,a,s_1)  \psi_d^{\dagger}(\p_2,x_2,b,s_2) \psi_d(\p_2-\Delta,x_2,b,s_2) \psi_u(\p_1+\Delta,x_1,a,s_1) |\text{Proton} \rangle \nonumber\\
&& \approx \int _{\k,\eta}
  \langle \text{Proton}|\sum_a\psi_u^{\dagger}(\p_1,x_1,a,s_1) \psi_u(\p_1+\Delta,x_1,a,s_1)|P,\k,\eta\rangle\langle P,\k,\eta| \sum_b \psi_d^{\dagger}(\p_2,x_2,b,s_2) \psi_d(\p_2-\Delta,x_2,b,s_2) |\text{Proton}\rangle, \nonumber\\
 &&= \langle \text{Proton}| \sum_a\psi_u^{\dagger}(\p_1,x_1,a,s_1) \psi_u(\p_1+\Delta,x_1,a,s_1)|P,\Delta,0\rangle\langle P,\Delta,0| \sum_b \psi_d^{\dagger}(\p_2,x_2,b,s_2) \psi_d(\p_2-\Delta,x_2,b,s_2) |\text{Proton}\rangle, \nonumber\\
  &&=4N_c^2 \mathcal{T}_u(x_1,0,\p_1,\Delta)\mathcal{T}_d^*( x_2,0,\p_2-\Delta,\Delta), 
  \end{eqnarray}
  where $\k$ and $\eta$ are the transverse and  longitudinal momentum transfer respectively, and the state $|P,\k, \eta\rangle$ denotes the proton state  with transverse momentum $\k$ and  longitudinal momentum $\bar P^-=(1+\eta)P^-$ with $P^-$  being the longitudinal momentum of the incoming proton. 
   In the above,  $\mathcal{T}_{u,d}$ is the unpolarized generalized transverse momentum dependent distribution (GTMD) \cite{gtmd1,gtmd2,gtmd3,gtmd4,gtmd5,gtmd6} defined as 
 \begin{equation}
\mathcal{T}_u(x,\eta,\p,\k)=\frac{1}{2N_c}\sum_{c,s}\langle \text{Proton}| \psi_u^{\dagger}(\p,x,c,s)\psi_u(\p+\k,x+\eta,c,s) |P,\k,\eta\rangle.   \label{gtmd}
 \end{equation}
 Note that for non-identical quark production, there is no longitudinal momentum transfer, i.e. $\eta=0$ in \eq{pd}.
 This is a direct consequence of the high energy eikonal approximation, where an energetic parton does not change its longitudinal momentum during the interaction with the target.

 Let us now consider the exchange contribution. We again insert the resolution of identity between the color singlet operators:
  \begin{eqnarray}\label{resolut}
 && \sum_{ab;s_1s_2} \langle \text{Proton}| \psi_u^{\dagger}(\p_1,x_1,a,s_1)  \psi_d^{\dagger}(\p_2,x_2,b,s_2) \psi_d(\p_2-\Delta,x_2,a,s_2) \psi_u(\p_1+\Delta,x_1,b,s_1) |\text{Proton} \rangle, \\
 &&= -\sum_i \sum_{s_1s_2}\langle \text{Proton}|\sum_{a} \psi_u^{\dagger}(\p_1,x_1,a,s_1)\psi_d(\p_2-\Delta,x_2,a,s_2)   |i\rangle\langle i|\sum_{b}\psi_d^{\dagger}(\p_2,x_2,b,s_2)\psi_u(\p_1+\Delta,x_1,b,s_1) |\text{Proton} \rangle, \nonumber
  \end{eqnarray}
  Here we have neglected the commutator term. This term is suppressed by one power of parton density, and is thus subleading at high energies, where the number of quarks in the proton is much larger than one. 
    
 This time obviously, single proton states do not contribute as the relevant matrix element vanishes. Instead  one expects the dominant contribution to arise due to neutron intermediate sates. We therefore write
 \begin{eqnarray}
  && \sum_{ab;s_1s_2} \langle \text{Proton}| \psi_u^{\dagger}(\p_1,x_1,a,s_1)  \psi_d^{\dagger}(\p_2,x_2,b,s_2) \psi_d(\p_2-\Delta,x_2,a,s_2) \psi_u(\p_1+\Delta,x_1,b,s_1) |\text{Proton} \rangle, \nonumber\\
 &=&-2N_c^2\mathcal{M}(x_1,\eta=x_2-x_1,\p_1,\p_2-\p_1-\Delta)\mathcal{M}^*( x_1,\eta=x_2-x_1,\p_1+\Delta,\p_2-\p_1-\Delta), 
 \end{eqnarray}
 where the off diagonal GTMD is defined as
  \begin{equation}
\mathcal{M}(x,\eta,\p,\k)=\frac{1}{2N_c}\sum_{c,s}\langle \text{Proton}| \psi_u^{\dagger}(\p,x,c,s)\psi_d(\p+\k,x+\eta,c,s) |N,\k,\eta\rangle.   \label{ogtmd}
 \end{equation}
 Here $|N,\k,\eta\rangle$ is the neutron state with the indicated transverse and longitudinal momentum. Within this approximation we thus have

\begin{eqnarray}
&&T_{ud}^D(p_1,p_2,\Delta)=4N_c^2 \mathcal{T}_u(x_1,\eta=0,\p_1,\Delta)\mathcal{T}_d^*( x_2,\eta=0,\p_2,\Delta),\nonumber\\
&& T_{ud}^E(p_1,p_2,\Delta)=-2N^2_c\mathcal{M}(x_1,\eta=x_2-x_1,\p_1,\p_2-\p_1-\Delta)\mathcal{M}^*( x_1,\eta=x_2-x_1,\p_1+\Delta,\p_2-\p_1-\Delta).
\end{eqnarray}
In all of the above expressions we are only considering unpolarized proton, and therefore we have assumed that all one particle averages have the structure $\delta_{s_1s_2}$ in spin indexes.

Substituting these expressions into \eq{IDE},  we finally obtain for production of
quarks of different flavors
\begin{eqnarray}\label{iud}
\mathcal{I}_{ud}&=&16(4N_c^2)\frac{(2\pi)^4}{4}\delta(p_1^{-}-q^-)\delta(p_1^{\prime -}-q^-)\delta(p_2^--q'^{-})\delta(p_2^{\prime -}-q'^{-})q^-q'^{-} \int_{p_1,p_2,\Delta} \nonumber\\
&\times&\Bigg\{
 \Big[\langle D(\p_1-\q+\Delta/2,2\Delta)D(\p_2-\q^\prime-\Delta/2,-2\Delta )\rangle -\frac{1}{N^2_c}\langle Q(\p_1-\q+\Delta/2,\p_2-\q^\prime-\Delta/2, \Delta)\rangle\Big]\nonumber\\
 &\times&\mathcal{T}_u(x_1,0,\p_1,\Delta)\mathcal{T}_d^*( x_2,0,\p_2,\Delta)\nonumber\\
 &-&\frac{1}{N_c}\Big[\langle Q(\p_1-\q+\Delta/2,\p_2-\q^\prime-\Delta/2, \Delta)\rangle-\langle D(\p_1-\q+\Delta/2,2\Delta)D(\p_2-\q^\prime-\Delta/2,-2\Delta )\rangle\Big]\nonumber\\
 &\times&\mathcal{M}(x_1,x_2-x_1,\p_1,\p_2-\p_1-\Delta)\mathcal{M}^*( x_1,x_2-x_1,\p_1+\Delta,\p_2-\p_1-\Delta)\Bigg\}.
\end{eqnarray}
The first term in this expression, up to a $1/N_c^2$ correction, describes independent scattering of the two quarks. The second term is quite interesting. It clearly describes correlate production and is suppressed by a single power of $1/N_c$. We will return to it after considering production of identical quarks.

 \subsection{Identical quarks}
 For identical quarks we have  to consider extra terms that arise due to antisymmetry of the amplitude $A$. 
 The antisymmetrization brings additional terms suppressed by $1/N_c$ due to the color index contraction structure in the two particle density operators. 
 The appropriate  form of the two particle densities in the proton state dominance approximation then is
\begin{eqnarray}
T_{uu}^D(p_1,p_2,\Delta)&=&4N_c^2 \mathcal{T}_u(x_1,\eta=0,\p_1,\Delta)\mathcal{T}_u^*( x_2,\eta=0,\p_2,\Delta)\nonumber\\
&-&2N_c\mathcal{T}_u(x_1,\eta=x_2-x_1,\p_1,\p_2-\p_1-\Delta)\mathcal{T}_u^*( x_1,\eta=x_2-x_1,\p_1+\Delta,\p_2-\p_1-\Delta), \nonumber\\
T_{uu}^E(p_1,p_2,\Delta)&=&4N_c\mathcal{T}_u(x_1,\eta=0,\p_1,\Delta)\mathcal{T}_u^*( x_2,\eta=0,\p_2,\Delta)\nonumber\\
&-&2N_c^2\mathcal{T}_u(x_1,\eta=x_2-x_1,\p_1,\p_2-\p_1-\Delta)\mathcal{T}_u^*( x_1,\eta=x_2-x_1,\p_1+\Delta,\p_2-\p_1-\Delta).\
\end{eqnarray}
 where $\mathcal{T}_u$ was defined in \eq{gtmd}.
Obviously, it is the proton and not the neutron intermediate state that now contributes both to the direct and the exchange term.
 
 Using the above relations, the expression for  production of identical quarks becomes 
\begin{eqnarray}\label{ifinal}
&&\mathcal{I}_{uu}= 16\frac{(2\pi)^4}{4}\delta(p_1^{-}-q^-)\delta(p_1^{\prime -}-q^-)\delta(p_2^--q'^{-})\delta(p_2^{\prime -}-q'^{-})q^-q'^{-} \nonumber\\
&&\times 2N^2_c\int_{p_1,p_2,\Delta}\Big[2 \langle D(\p_1-\q+\Delta/2,2\Delta)D(\p_2-\q^\prime-\Delta/2,-2\Delta )\rangle \mathcal{T}_u(x_1,0,\p_1,\Delta)\mathcal{T}^*_u( x_2,0,\p_2,\Delta)   \nonumber\\
&&-\frac{1}{N_c}\langle Q(\p_1-\q+\Delta/2,\p_2-\q^\prime-\Delta/2, \Delta)\rangle \mathcal{T}_u(x_1,x_2-x_1,\p_1,\p_2-\p_1-\Delta)\mathcal{T}^*_u( x_1,x_2-x_1,\p_1+\Delta,\p_2-\p_1-\Delta)\Big].\nonumber\\
\end{eqnarray}
Just like for nonidentical quarks, this expression contains $1/N_c$ suppressed terms. 
Indeed one expects such terms to be present due to quantum statistics effects. In the next section we show that this is  the correct physical interpretation of the $1/N_c$ suppressed terms in eq.(\ref{ifinal}).

\section{Quadrupole, Pauli Blocking, HBT and all that}

It is instructive to see explicitly how the two types of effects due to quantum statistics discussed in the literature, namely the Pauli blocking in the projectile wave function and the Hanbury Brown-Twiss (HBT) correlations appear from different kinematic regions of the quadrupole contribution in eq.(\ref{ifinal}).

In order to make this discussion simple we start by considering a simplified limit of the quadrupole correlator corresponding to the week target field limit. In other words, we expand the quadrupole to the lowest nontrivial order in the target field, and estimate the correlator utilizing the simple Gaussian field ensemble. Later in this section we demonstrate that  the weak field limit has in fact the same structure as the complete expression and the identification of the quantum statistics effects is not limited to the weak field limit.

The lowest order term in the expansion of the quadrupole operator that is relevant to our discussion is the one where each eikonal scattering matrix is expanded to first order in the field:
\begin{equation}
Q(\x_1,\x_2,\x_3.\x_4)\approx \frac{1}{N_c}\alpha^a(\x_1)\alpha^b(\x_2)\alpha^c(\x_3)\alpha^d(\x_4)Tr[\tau^a\tau^b\tau^c\tau^d]. 
\end{equation}
Here $\alpha^a(\x)$ is the phase of the eikonal scattering matrix $U(\x)=\exp\{i\alpha^a(\x)\tau^a\}$ with $\tau^a$ being the generator of $SU(N_c)$ in the fundamental representation. 
We now employ simple Gaussian averaging model for averaging over the target with the basic contraction \cite{mv}, 
\begin{equation}
\langle \alpha^a(\x)\alpha^b(\y)\rangle=\delta^{ab}\mu^2(\x-\y).
\end{equation}
 Note that in Fourier space  $\mu^2(\p)$ has the interpretation of the target gluon GTMD. The assumption of the translational invariance of the target in this language translates into the statement that the GTMD has support only for vanishing momentum transfer. Also by assumption, the target GTMD does not depend on the longitudinal momentum fraction. We will thus write in momentum space 
\begin{equation}
\langle \alpha^a(\k_1)\alpha^b(\k_2)\rangle=\delta^{ab}\frac{2}{N_c}G_T(\k_1)\delta^2(\k_1-\k_2). 
\end{equation}
where
 \begin{equation}
G_T(\k)\equiv \frac{N_c}{2}\int d^2\x e^{i\k\x}\mu^2(\x). 
\end{equation}
Within this Gaussian model, the average of the quadrupole in the large $N_c$ limit is given by two contractions
  \begin{equation}\label{factorq}
Q(\x_1,\x_2,\x_3.\x_4)\approx\frac{N_c^2}{4}\Big[\mu^2(\x_1-\x_2)\mu^2(\x_3-\x_4)+\mu^2(\x_1-\x_4)\mu^2(\x_2-\x_3)\Big]. 
\end{equation}
As we show now, the first one of this terms naturally corresponds with the Pauli blocking effect in the proton wave function, while the second one with the HBT effect in quark production. We will then explain that the exact quadrupole amplitude does indeed approximately factorize in a way similar to Eq.\,(\ref{factorq}) in large parts of the phase space. The quantum statistics effects are not arbitrarily introduced by the approximation Eq.\,(\ref{factorq}), but rather  arise from parts of the phase space integral  where this factorization is a good approximation to the quadrupole amplitude.

\subsection{The Pauli blocking}
Remembering  our definition of the Fourier transform eq.(\ref{ft}) and assignments of transverse momenta, we find that the  Fourier transform of the first term in Eq.\,(\ref{factorq}) is 
\begin{equation}\label{pbc}
S\delta^2(\Delta)G_T(\p_1-\q)G_T(\p_2-\q'), 
\end{equation}
where $S$ is the total area of the target. The corresponding contribution to the production cross section in Eq.\,(\ref{ifinal}) is (we drop trivial kinematic factors for simplicity) 
\begin{equation}\label{pb}
\mathcal{I}_{PB}\propto -\frac{2}{N_c}S\int_{\p_1,\p_2}G_T(\p_1-\q)G_T(\p_2-\q') |\mathcal{T}_u(x_1,x_2-x_1,\p_1,\p_2-\p_1)|^2.
\end{equation}
To elucidate the meaning of this term further, let us take (for illustration only) a slight generalization of the simple ansatz used frequently for the momentum transfer dependence of the GTMD
\begin{equation}\label{tmd}
\mathcal{T}_u(x_1,x_2-x_1,\p_1,\p_2-\p_1)=T_u(x_1,\p_1)f(x_1-x_2)\mathcal{F}(\p_1-\p_2), 
\end{equation}
where $T_u(x,\p)$ is the conventional parton transverse momentum dependent (TMD), while the dependence on the momentum transfer is contained in the form factor \cite{kr1}, 
\begin{equation}\label{f}
\mathcal{F}(\p)=\frac{1}{(\p/\Lambda)^2+1}, 
\end{equation}
where the nonperturbative scale $\Lambda$ roughly has the meaning of the inverse proton radius.

 For completeness we have also introduced a longitudinal form factor $f$, which should be maximal at $x_1-x_2=0$ and presumably drops for large differences between the momentum fractions. This longitudinal from factor is not important for our discussion.
 

It is now obvious that Eq.\,(\ref{pb}) has exactly the Pauli blocking form discussed in detail in Ref.\,\cite{paulib}. The difference between the momenta of the two {\bf incoming} quarks is pinned to be smaller than the infrared scale $\Lambda$. Formally the translationally invariant projectile considered in Ref.\,\cite{paulib} corresponds to the limit  $\Lambda\rightarrow 0$. In this limit the two incoming momenta are exactly equal. This particular configuration in the incoming proton wave function is suppressed due to the Pauli blocking effect (note the negative sign in Eq.\,(\ref{pb})). The two quarks then scatter independently off the target with the appropriate momentum transfer to produce the final state quarks. For non-vanishing $\Lambda$  the suppression is maximal at $\p_1=\p_2$, but is not  restricted to exactly equal momenta. Rather it is effective within a small nonperturbative width $\Lambda$. Clearly these general features do not depend neither on the exact form of the form factor Eq.\,(\ref{f}), nor on the factorizable ansatz Eq.\,(\ref{tmd}). 

\subsection{The HBT correlations}
Consider now the second term in Eq.\,(\ref{factorq}). This term has the following structure, 
\begin{equation}\label{hbtg}
S\delta^2(\p_1-\p_2+\Delta-\q+\q')G_T(\p_1-\q)G_T(\p_2-\q'). 
\end{equation}
The corresponding contribution to particle production becomes (again dropping the irrelevant kinematic factors)
\begin{equation}\label{hbt}
\mathcal{I}_{HBT}\propto -\frac{2}{N_c}S\int_{\p_1,\p_2}G_T(\p_1-\q)G_T(\p_2-\q') \mathcal{T}_u(x_1,x_2-x_1,\p_1,\q'-\q)\mathcal{T}^*_u(x_1,x_2-x_1,\p_2+\q-\q',\q'-\q), 
\end{equation}
Again, using the ansatz of Eqs.\,(\ref{tmd}, \ref{f}) makes the physical meaning of this term transparent. It is clear that the above term has the typical form of the HBT correlation. This time the difference between the momenta of the {\bf final state} particles is pinned to within the nonperturbative scale $\Lambda$, which has the meaning of the proton radius.
In the extreme limit $\Lambda\rightarrow 0$, the two outgoing momenta are exactly equal, while the rest of the expression gives the probability of the two incoming particles with equal momenta $\p_1$ to scatter into the final state with momentum $\q$. The width of the correlated peak is directly determined by the nonperturbative scale $\Lambda$. This is different from the Pauli blocking effect discussed in the previous subsection, where the width of the trough in the observed momenta is affected by the momentum transfer from the  target.

\subsection{No need in Gaussian factorization assumption}
In the two previous subsections  we have employed the Gaussian factorization model in order to see explicitly the main effects in a simple manner. One might therefore wonder if our conclusions hinge crucially on this factorization hypothesis. Here we wish to explain that this is not the case, and in fact the factorization  is a reasonable proxy to the properties of the quadrupole in large part of the phase space which is integrated over to calculate the complete cross-section.

To understand this, consider the calculation of the quadrupole correlator in an ensemble of target fields. We assume that the target is characterized by some saturation momentum $Q_s$. It is known that the saturation momentum plays the role of the inverse correlation length of the target fields.  Let us examine the quadrupole correlator for points such that $|r|=|\x_1-\x_2|\sim |r'|=|\x_3-\x_4|\sim 1/Q_s$, but $|R|\sim|\x_1-\x_3|\gg 1/Q_s$. Our target field ensemble is such that in this regime the fields at point $\x_1$ is correlated with the fields at point $\x_2$, and the fields at $\x_3$ are correlated with the fields at $\x_4$. However the points in the vicinity of $\x_1$ and $\x_2$ are completely uncorrelated with the points in the vicinity of $\x_3$ and $\x_4$. Thus in this regime we must have
\begin{equation}
\lim_{R\sim |x_1-x_3|\gg1/Q_s}Q(\x_1,\x_2,\x_3,\x_4)= \frac{1}{N_c}\langle U(\x_1)_{ab}U^\dagger(\x_2)_{bc}\rangle\langle U(\x_3)_{cd}U^\dagger(\x_4)_{da}\rangle=D(\x_1-\x_2)D(\x_3-\x_4).
\end{equation}
Here we have assumed that the color neutralization of the target ensemble occurs on distance scales $\sim 1/Q_s$, and therefore each one of the averages separately is color invariant, i.e.
\begin{equation}
\langle U(\x_1)_{ab}U^\dagger(\x_2)_{bc}\rangle=\delta_{ac}\frac{1}{N_c}\langle Tr[U(\x_1)U^\dagger(\x_2)]\rangle.
\end{equation}
For these large $R$, integration over $R$ in the calculation of cross-section results in the factor $\delta(\Delta)$, precisely as in Eq.\,(\ref{pbc}). Thus this region of the phase space quite generally leads to the Pauli blocking contribution without the need for technical assumption of Gaussian factorization.

Similar  argument for the structure of the quadrupole correlator holds in the part of the phase space where $|\x_1-\x_4|\sim |\x_2-\x_3|\sim 1/Q_s$, but $|\x_1-\x_2|\gg 1/Q_s$. In this case we have 
\begin{equation}
\lim_{|x_1-x_2|\gg1/Q_s} Q(\x_1,\x_2,\x_3,\x_4)=\frac{1}{N_c}\langle U^\dagger(\x_4)_{da} U(\x_1)_{ab}\rangle\langle U^\dagger(\x_2)_{bc}U(\x_3)_{cd}\rangle=D(\x_1-\x_4)D(\x_2-\x_3).
\end{equation}
Now integration over $\x_1-\x_3$ results in $\delta^2(\p_1-\p_2+\Delta-\q+\q')$ as in Eq.\,(\ref{hbtg}), again obviating the redundancy of the Gaussian factorization hypothesis.

We thus conclude that the general structure of the quadrupole correlator ensures the presence of the quantum statistics effects  in the form presented above, in Eqs.\,(\ref{pb},\ref{hbt}) with the target TMD's $G_T$ defined as the Fourier transform of the dipole amplitude:
 \begin{equation}
G_T(\k)\equiv \int d^2\x e^{i\k\x}D(\x). 
\end{equation}

\subsection{Other correlations}
The quantum statistics effects we discussed above, clearly lead to correlations between the emitted quarks. In principle this is not the only source of the correlated production.  The first term in Eq.\,(\ref{ifinal}) is  equal to the square of the single inclusive  production probability only if one approximates the average of the product of the two dipoles by the product of averages. Any deviations from this independent scattering approximation produce correlations in the spectrum. However, these correlations at least naively are suppressed at large $N_c$ relative to the ones associated with quantum statistics, since the corrections to the independent scattering are generically of order $1/N_c^2$.

For example in the ``glasma graph'' approximation, which amounts to expanding the product of dipoles to the same order as we did for the quadrupole above, one obtains \cite{ridge1,bapp}
\begin{equation}\label{glgr}
\langle D(\x_1,\x_2)D(\x_3,\x_4)\rangle\approx\langle D(\x_1,\x_2)\rangle\langle D(\x_3,\x_4)\rangle +\frac{1}{N_c^2-1}\Big[\langle D(\x_1,\x_3)\rangle\langle D(\x_2,\x_4)\rangle+\langle D(\x_1,\x_4)\rangle\langle D(\x_2,\x_3)\rangle\Big].
\end{equation}
In fact part of this expression  is quite generic, even though it has been derived within the glasma graph approach. It has a natural interpretation along the lines of the previous subsection. The last term in Eq.\, (\ref{glgr}) in physical  terms should be read as \footnote{The second term on the RHS of Eq.\,(\ref{glgr}) does not have the same generality. Assuming the invariance of the target under the diagonal $SU(N)$ group, instead of the product of two dipoles we get terms of the type $\langle Tr[U(x_1)]Tr[U(x_3)]\rangle \langle Tr[U^\dagger(x_2)]Tr[U^\dagger(x_4)]\rangle$ and  $\langle Tr[U(x_1)U(x_3)]\rangle \langle Tr[U^\dagger(x_2)U^\dagger(x_4)]\rangle$. This can only be written as a product  of dipole amplitudes (as in  Eq.\,(\ref{glgr}) in the lowest order in the expansion in the target fields. In the dense target regime this term is not only suppressed by a factor proportional to $1/N_c^2$, but also vanishes much faster than the product of two dipoles as a function of  the intensity of the target field \cite{gf}.}
\begin{equation}
\lim_{ |\x_1-\x_4|\sim 1/Q_s; \  |x_2-\x_3|\sim 1/Q_s;\  |\x_1-\x_2|\gg 1/Q_s}\langle D(\x_1,\x_2)D(\x_3,\x_4)=\frac{1}{N_c^2-1}\langle D(\x_1,\x_4)\rangle\langle D(\x_2,\x_3)\rangle.
\end{equation}  
Thus even though this type of term produces correlations from the integral over this region of phase space,  the magnitude of these correlations is suppressed by one additional power of $1/N_c$ with respect to the quantum statistics effects in the last term in Eq.\,(\ref{ifinal}). 
 
 This is not to say that all the correlated production effects stemming from the two dipole  term in eq.(\ref{ifinal})  are suppressed as $1/N_c^2$. In particular, there is a contribution from the part of the phase space where all the coordinates $\x_i$ are within the distance $1/Q_s$ of each other. For such configurations $\langle D(\x_1,\x_2)D(\x_3,\x_4)\rangle\ne \langle D(\x_1,\x_2)\rangle\langle D(\x_3,\x_4)\rangle$, and thus some correlated production arises. This part of the phase space integration must include effects due to density variation and local anisotropy of the target fields, discussed in Refs.\,\cite{ridge3,ridge2}. 
These effects are not suppressed by powers of $1/N_c$ relative to the independent production, and are thus enhanced by one power of $N_c$ relative to the effects due to quantum statistics. There is however a price to pay. Since this contribution arises from a small region of  space, where all the points $\x_i$ are close to each other, it is suppressed by a factor $1/S$, where $S$ is the area of the target relative to the correlated quadrupole contribution. The actual suppression factor is of course dimensionless, and is actually $1/Sq^2$ for large transverse momentum $q^2>Q_s^2$. Thus the ratio between the two correlated contributions is roughly $N_c/q^2S$, which is likely to lean in favor of the quantum statistics effects already at moderate transverse momenta. Note that the signs of these two correlated contributions are opposite, in the sense that the quantum statistics suppresses production of two quarks with same momentum, while the effect of the two dipole term in fact enhances this  production. 

It is therefore our conclusion that in the framework of the CGC calculations the quantum statistics effects are parametrically the leading source for correlated quark-quark production.

\subsection{Back to non identical particles}
To complete our discussion of quantum statistics we would like to compare eq.(\ref{ifinal}) with the similar expression for non identical particle production eq.(\ref{iud}). 

The first term in Eq.(\ref{iud}) up to a $1/N_c^2$ correction corresponds to an independent scattering of the $u$ and $d$ quarks drawn from the proton wave function. 

The second term is more interesting and it does yield a correction which must contain correlations. First thing to note about this term is that, as expected it does not contain any analog of the Pauli blocking. Recall that the Pauli blocking contribution arises from the part of the integration space where $Q(\x_1,\x_2,\x_3,\x_4)=D(\x_1,\x_2)D(\x_3,\x_4)$. Obviously this contribution cancels in the second term in eq.(\ref{iud}) between the quadrupole and the two dipole terms.

On the other hand eq.(\ref{iud}) does seem to contain a correlated contribution from the integration region which for identical quarks yields the HBT correlation, namely where  $Q(\x_1,\x_2,\x_3,\x_4)=D(\x_1,\x_4)D(\x_2,\x_3)$. The interpretation of  this contribution in eq.(\ref{ifinal}) is fairly straightforward. The two quarks with momenta $\p_1$ and $\p_2$ are drawn independently from the proton wave function with the probability proportional to the product of particle densities. They then scatter independently, but the outgoing quarks destructively interfere if their final momenta are similar.

For non identical quarks the interference is of course not possible, and thus the interpretation of this correlated contribution is not straightforward. The magnitude of this contribution is not proportional to product of particle densities, but rather to the product of the off diagonal proton-neutron matrix elements $\mathcal{M}$. The internal structures of the proton and the neutron are quite similar, and thus naively one does not expect that the factor $\mathcal{M}\mathcal{M}^*$ in eq.(\ref{iud})  is qualitatively different from the factor $\mathcal{T}\mathcal{T}^*$ in eq.(\ref{ifinal}).

There is a difference between the two contributions that may turn out to be significant. The operator $\psi^\dagger_u\psi_u$ is a linear combination of an isoscalar and  and an isovector operators  $\psi^\dagger_u\psi_u=I_0+I^0_1$
with $I_0=\frac{1}{2}(\psi^\dagger_u\psi_u+\psi^\dagger_d\psi_d)$ and  $I^0_1=\frac{1}{2}(\psi^\dagger_u\psi_u-\psi^\dagger_d\psi_d)$, while the operator   $\psi^\dagger_u\psi_d=I^+_1$ is a pure isovector. It may be natural to associate the HBT correlation mostly with the matrix element of the isosinglet operator $I_0$. It  is thus possible that the isosinglet proton matrix element $\langle P|I_0|P\rangle$ is numerically larger than the isovector one $\langle P|I^0_1|P\rangle$. Since by Wigner-Eckart theorem 
$\langle P|I^0_1|P\rangle=\frac{1}{2}\langle P|I^+_1|N\rangle$, the correlated contribution to non identical quarks may turn out to be significantly smaller than the HBT effect for identical particles.
 Unfortunately one cannot make a firm  statement, since both these quantities are determined by strong nonperturbative physics.
 
 Nevertheless it is interesting that our calculation does point to an existence of correlations  for nonidentical particles which are in a certain sense analogous to the HBT effect.

\section{Discussion}
To summarize we would like to make some concluding remarks. The first point we would like to make is that although our calculation is performed in the forward kinematics, there are many analogies with CGC calculational approach at central rapidities. In particular our calculation is very similar to the calculation of the Pauli blocking effect of Ref.\, \cite{paulib}. The main difference is that our present calculation is much simpler and therefore makes the interested effects be easier to identify. In Ref.\,\cite{paulib} the origin of quarks in the wave function was specifically assumed to be due to splitting of gluons. This is of course how things are at high energy far enough from the forward (proton moving) direction.  This however complicates the calculation due to explicit presence of the gluon emitter in the wave function. Thus the color structures appearing in Ref.\,\cite{paulib} are somewhat more complicated and to some degree obscure the simplicity of the terms responsible for quantum statistics effects. 

The most significant difference between our calculation and calculation of double inclusive production at central rapidity is, of course that in the latter the contribution of quarks is negligible, and the main source of particle production is scattering of gluons. But forgetting for a moment about the identity of scattered partons, the analogy even here  is rather close. The 2GTMD that arises in our calculation, at central rapidities  takes on the identity of the average of the four gluon operators in the projectile wave function, or equivalently four powers of the projectile charge density. The factorizable approximation discussed in Sec. III is directly analogous to calculation of this average in the McLerran-Venugopalan model, where the four gluon average factorizes into products of two two- gluon functions. The factorization itself is somewhat  different, since for gluonic operators one can have an additional non-vanishing average, $\langle a^\dagger a^\dagger\rangle$, which cannot appear for quarks due to baryon number and charge conservation. Nevertheless, the structure of the two calculations is quite similar. It would be therefore interesting to analyze the CGC calculation from the point of view of the current paper. It seems very likely to us that the main origin of the correlated production in that case is again the quadrupole amplitude (this time adjoint quadrupole) and not the two dipole term.

Our second comment is related to particle correlations stemming from the Pauli blocking contribution effect. As discussed previously (Refs.\,\cite{bosee,paulib}), this type of correlation in the projectile wave function is reflected in the momentum distribution of produced particles if momentum transfer due to scattering is not too large, so that the momenta of produced particles are not too different from the momenta of the particles in the projectile wave function. This regime is strictly speaking outside of the validity of our calculation, since we have assumed that most of the transverse momentum  of  produced particles originates in the momentum transfer from the target. If this is not the case, the calculation ceases to be manageable since the spectator effects become significant, and those cannot be accounted for without the knowledge of the nonperturbative proton wave function (see Appendix). Although we believe that qualitatively the effect of the spectators should not be too significant at large transverse momentum, one has to bear in mind this limitation of our calculation. 

Finally we note that the calculation presented here, albeit very simple may have direct relevance to phenomenology. Although fragmentation and  hadronization  will undoubtedly broaden the minimum in the correlation function produced by the parton level quantum statistics effects, the basic signature is likely to remain visible in physical observables. The most natural application of these ideas would be in high $p_T$ forward dijet production. The typical momentum transfer from the nuclear target is of order $Q_s$. Nevertheless the momentum distribution  beyond $Q_s$ falls only as a power and the production of  hard jets in $p-A$ scattering  is not suppressed. The naive expectation is that roughly half of the dijet events at forward rapidity should originate from identical $u$ quarks in the proton. Thus quantum statistics effects should be quite significant for this observable. The same goes for double inclusive hadron production. One expects for example that these effects will be stronger in $\pi^+\pi^+$ and $\pi^-\pi^-$ production than in $\pi^+\pi^-$ production, since in the former case a larger proportion of produced hadrons should come from fragmentation of identical quarks. Similarly these effects must be stronger in baryon-baryon production than in baryon-  
antibaryon production for the same reason. As suggested in \cite{paulib}, this may be a candidate explanation for the ALICE observation of depletion of the same side production of baryon-baryon and 
antibaryon - antibaryon pairs, and absence of such effect for 
baryon-antibaryon  pairs \cite{baryonpairs}.

\section{Appendix}

In this appendix we discuss some problems faced by any parton model like calculation of particle production in QCD. This discussion is of course not new, and in particular has large overlap with Refs.\, \cite{urs,cut} but we feel it may be helpful to make this point explicitly in relation to our calculation as well as other recent calculations in the literature. We also point out that the calculations at central rapidity are in a somewhat better  position in this respect. 

Although we are interested in double inclusive particle production, for simplicity let us concentrate on a single particle production, with extension to other observables being obvious. Say we want to calculate single inclusive production of quarks in the final state.  Let us for a second forget about the problem of confinement, and assume that one can actually produce free quarks in the final state, and think of the proton as a  bound state of quarks. The incoming proton state $|P\rangle$ gives rise to the outgoing state
\begin{equation}
|out\rangle=\hat S|P\rangle, 
\end{equation}
where $\hat S$ is the S-matrix operator. The simplest observable we can form to measure the quark number in the outgoing state is
\begin{equation}\label{qq}
N_{out}(k)=\langle out|q^\dagger(k) q(k)|out\rangle\equiv\sum_X\langle out|k,X\rangle\langle k,X|out\rangle, 
\end{equation}
where the summation is over all the states of ``spectator'' degrees of freedom. 
This is however clearly not what one is interested in. This observable measures the quark number in any outgoing state, even if it is simply an elastically scattered proton. The wave function of such a bound state has non-vanishing overlap with quark Fock space states, and thus will give a contribution to Eq.\,(\ref{qq}). Since we are only interested in production of quarks which are not part of a scattered proton, or of any other bound state for that matter, we must at least subtract elastically scattered proton states from the outgoing state before measuring the quark content. Let us define
\begin{equation}
|\bar {out}\rangle=|out\rangle-\sum_i |P_i\rangle\langle P_i|out\rangle, 
\end{equation}
where  the index $i$ indicates all possible internal quantum numbers of the proton stat, such as its transverse momentum, spin orientation or more generally internal excitation if it corresponds to another stable bound state. 
The appropriate observable is then
\begin{eqnarray}\label{n}
N(k)=\sum_X\langle \bar{out}|k,X\rangle\langle k,X|\bar{out}\rangle&=&\sum_X\Bigg[\langle out|k,X\rangle\langle k,X|out\rangle+\langle out|P_i\rangle\langle P_i|k,X\rangle\langle k,X|P_j\rangle\langle P_j|out\rangle\nonumber\\
&-&\langle out|P_i\rangle\langle P_i|k,X\rangle\langle k,X|out\rangle-\langle out|k,X\rangle\langle k,X|P_j\rangle\langle P_j|out\rangle\Bigg]. 
\end{eqnarray}
This expression is  rather complicated, since it involves the overlap between the outgoing state and a state of elastically scattered proton. In the language of the amplitudes $A$ introduced in the text this overlap involves expressions of the type $\sum_XA(k,X)A^*(k,SX)$, where $SX$ is the configuration of spectator partons after scattering. Such matrix elements depend on the target on which the initial parton scatters. They are thus non universal and unlike 2GTMD's cannot be used o parametrized the proton structure.
Through these amplitudes the scattering of spectator partons affects the  single inclusive production, and clearly also the double inclusive production.
  
In QCD where there is confinement and free quarks cannot be produced, the situation is even more complicated, since there is an ambiguity as to what states exactly have to be subtracted. The only way to resolve this, is to consider directly production of mesons, but that of course requires  additional nonperturbative information about the wave function of mesons.

The one redeeming feature of Eq.\,(\ref{n}), is that all the terms but the first one involve direct overlap of the proton wave function with a quark at the observed momentum. If we measure quarks at large transverse momentum which is not contained in the proton, this overlap vanishes and one neglect all the terms except for the first one. In this parton model like approximation Eq.\,(\ref{n}) reduces to Eq.\,(\ref{qq}). This is the expression  used in the present paper. 

We stress again that its use requires that most of the transverse momentum of produced particles originates from the momentum transfer with the target and not from the wave function of the incoming proton. As noted in the text, the correlations generated by the Pauli blocking contribution require the opposite momentum balance, and thus are affected by the spectator scattering contributions.

We note that the situation is a little different in the CGC calculational scheme used to calculate particle production at mid rapidity. In this calculation, as explained in Refs.\,\cite{urs,cut} one does not calculate the number of  bare quarks (and/or gluons) in the final state, but redefines the quark operator so that it carries all its Weiszacker-Williams cloud of gluons. As a result such an operator has vanishing overlap (in perturbation theory) with any relevant state. Thus at least formally one does not neglect the analog of the three last terms in Eq.\,(\ref{n}) but takes them into account perturbatively constructing the proton wave function at low $x$.


\begin{acknowledgments}
 A.K. would like to thank the Particles and Nuclear Physics Group of  the Universidad Santa Maria for the hospitality while this work was being done. This research was supported in part by Fondecyt grant 1150135, ECOS-Conicyt C14E01, Anillo ACT1406 and Conicyt PIA/Basal FB0821 (A.R.) and  the NSF Nuclear Theory grant 1614640 and  Conicyt (MEC) grant PAI 80160015 (A.K.).
\end{acknowledgments}



\begin{thebibliography}{99}

\bibitem{exp-r1}
V. Khachatryan {\it et al.} (CMS Collaboration), JHEP {\bf 09}, 091 (2010) [arXiv:1009.4122]; Phys. Lett. {\bf B718}, 795 (2013) [arXiv:1210.5482]; Phys. Lett. {\bf B724}, 213 (2013).
\bibitem{exp-r2}
 B. Abelev  {\it et al.}  (ALICE Collaboration), Phys. Lett. {\bf B719}, 29 (2013) [arXiv:1212.2001]; Phys. Lett. {\bf B726}, 164 (2013) [arXiv:1307.3237];
 Phys. Rev. {\bf C90}, 054901 (2014). 
\bibitem{exp-r3}
G. Aad  {\it et al.}  (ATLAS Collaboration), Phys. Rev. Lett. {\bf 110}, 182302 (2013) [arXiv:1212.5198]; Phys. Lett. B {\bf 725}, 60 (2013); ATLAS-CONF-2014-021. 

\bibitem{exp-r4}
 A. Adare {\it et al.} (PHENIX Collaboration), Phys. Rev. Lett. {\bf 111}, 212301 (2013) [arXiv:1303.1794]; Phys. Rev. Lett. {\bf 114}, 192301 (2015) [arXiv:1404.7461].

\bibitem{exp-r5}
L. Adamczyk {\it et al.} (STAR Collaboration), Phys. Lett. {\bf B743}, 333 (2015) [arXiv:1412.8437]; Phys. Lett. {\bf B747}, 265 (2015) [arXiv:1502.07652]. 

\bibitem{hydro} 
P. Bozek, Eur. Phys. J. {\bf C71}, 1530 (2011) [1010.0405]; Phys. Rev. {\bf C88}, 014903 (2013); P. Bozek and W. Broniowski, Phys.
Lett. {\bf B718}, 1557 (2013) [arXiv:1211.0845].

\bibitem{ridge0}
A. Dumitru, F. Gelis, L. McLerran, and R. Venugopalan, Nucl. Phys. {\bf A810}, 91 (2008) [arXiv:0804.3858]; S. Gavin, L. McLerran, and G. Moschelli,
Phys. Rev. {\bf C79}, 051902  (2009) [arXiv:0806.4718]; Y. V. Kovchegov, E. Levin, and L. D. McLerran, Phys. Rev. {\bf C63}, 024903 (2001) [hep-ph/9912367].
\bibitem{ridge1}
A. Dumitru, K. Dusling, F. Gelis, J. Jalilian-Marian, T. Lappi and Venugopalan  Phys. Lett. {\bf B697}, 21 (2011) [arXiv:1009.5295].
\bibitem{ridge2}
A. Kovner and M. Lublinsky, Phys. Rev. {\bf D83}, 034017 (2011) [arXiv:1012.3398]; Phys. Rev. {\bf D84}, 094011 (2011) [arXiv:1109.0347].
\bibitem{ridge3}
E. Levin and A. H. Rezaeian, Phys. Rev. {\bf D84}, 034031 (2011) [arXiv:1105.3275]; E. Iancu and A. H. Rezaeian, Phys. Rev. {\bf D95}, 094003 (2017) [arXiv:1702.03943]. 
\bibitem{ridge1-raju}
K. Dusling and R. Venugopalan, Phys. Rev. Lett. {\bf 108}, 262001 (2012); Phys. Rev. {\bf D87}, 094034 (2013) [arXiv:1302.7018].
\bibitem{ridge33}
E. Iancu and D. Triantafyllopoulos, JHEP {\bf 1111}  105 (2011) [arXiv:1109.0302].
\bibitem{ridge4}
 Y.~V.~Kovchegov and D.~E.~Wertepny, Nucl. Phys. {\bf A925}, 254  (2014) [arXiv:1310.6701]. 
 
\bibitem{ridge5}
For a review see: A. Kovner and M. Lublinsky, Int. J. Mod. Phys. {\bf E22}, 1330001 (2013) [arXiv:1211.1928] and references
therein.




\bibitem{bosee}  
N. Armesto, T. Altinoluk, G. Beuf, A. Kovner and M. Lublinsky, Phys. Lett. {\bf B751} (2015) 448 [arXiv:1503.07126 ]. 


\bibitem{hbthadrons}  
Y. V. Kovchegov and D. E. Wertepny, Nucl. Phys. {\bf A906}, 50 (2013) [arXiv:1212.1195]; N. Armesto, T. Altinoluk, G. Beuf, A. Kovner and M. Lublinsky, Phys. Lett. {\bf B752}, 113 (2016) [arXiv:1509.03223];
 E. Gotsman, E. Levin and U. Maor,  Eur.Phys. J. {\bf C76}, 607 (2016) [arXiv:1607.00594]; E. Gotsman and E. Levin,  Phys.Rev. {\bf D95}, 014034 (2017) [arXiv:1611.01653], E. Gotsman and E. Levin, arXiv:1705.07406.  
 

\bibitem{paulib}  
N. Armesto, T. Altinoluk, G. Beuf, A. Kovner and M. Lublinsky, Phys. Rev. {\bf D95}  034025 (2017) [arXiv:1610.03020].  


\bibitem{2qg} A. Kovner, M. Lublinsky and V. Skokov, arXiv:1706.02330. 
 
 \bibitem{kr1} 
 A. Kovner and A. H. Rezaeian, Phys. Rev. {\bf D95}, 114028 (2017) [arXiv:1701.00494].
 
 \bibitem{kr} 
A. Kovner and A. H. Rezaeian, Phys. Rev. {\bf D90}, 014031 (2014) [arXiv:1404.5632];
Phys. Rev. {\bf D92}, 074045(2015) [arXiv:1508.02412].  


\bibitem{dj-rhic}
A. Dumitru, A. Hayashigaki and J. Jalilian-Marian,  
  Nucl.\ Phys.\  {\bf A765}, 464 (2006) [hep-ph/0506308]; Nucl. Phys. {\bf A770}, 57 (2006) [hep-ph/0512129].
  
  
 \bibitem{jimwlk}
J. Jalilian-Marian, A. Kovner, A. Leonidov and H. Weigert, Nucl. Phys. {\bf B504}, 415 (1997); {\it ibid.}, Phys. Rev. {\bf D59}, 014014
(1999); E. Iancu, A. Leonidov and L. D. McLerran, Nucl. Phys. {\bf A692}, 583 (2001); E. Ferreiro, E. Iancu, A. Leonidov and L. D. McLerran, Nucl. Phys. {\bf A703}, 489 (2002). 
\bibitem{bk}
I.~Balitsky,
Nucl.\ Phys.\  {\bf B463}, 99 (1996);
Y.~V.~Kovchegov,
Phys.\ Rev.\   {\bf D60}, 034008 (1999);
Phys.\ Rev.\   {\bf D61}, 074018 (2000).

\bibitem{marke}
B. Blok, Y. Dokshitzer, L. Frankfurt and M. Strikmam, Phys. Rev. {\bf D83}, 071501 (2011). 

\bibitem{2gpd-1}
A. Del Fabbro and D. Treleani, Phys. Rev. {\bf D61}, 077502 (2000). 
\bibitem{2gpd-2}
M. Diehl, Eur. Phys. J. {\bf C25}, 223 (2002); Erratum-ibid. {\bf C31},277 (2003);
Phys. Rept. {\bf 388}, 41 (2003). 
\bibitem{2gpd-3}
J. R. Gaunt and W. J. Stirling, JHEP {\bf1003}, 005 (2010) [arXiv:0910.4347]. 
\bibitem{2gpd-4}
M. Diehl, D. Ostermeier and A. Schafer, JHEP {\bf 1203}, 089  (2012) [arXiv:1111.0910]. 

\bibitem{2gpd-5}
K. Golec-Biernat and A. M. Stasto, Phys. Rev. {\bf D95}, 034033 (2017) [arXiv:1611.02033]. 

\bibitem{dipoles}
T. Lappi, B. Schenke, S. Schlichting and R. Venugopalan, JHEP {\bf 1601}, 061 (2016) [arXiv:1509.03499]; K. Dusling, M. Mace and R. Venugopalan,  arXiv:1706.06260; arXiv:1705.00745. 

\bibitem{vladi} 
  A.~Dumitru, L.~McLerran and V.~Skokov,
  Phys.\ Lett.\  {\bf B743}, 134 (2015)
  [arXiv:1410.4844];
     A.~Dumitru and V.~Skokov,
  arXiv:1411.6630 [hep-ph];
  V.~Skokov,
  arXiv:1412.5191 [hep-ph]. 

\bibitem{gf}
F.~Gelis and J.~Jalilian-Marian,
  Phys.\ Rev.\  {\bf D66}, 014021 (2002)
  [hep-ph/0205037]; Phys. Rev. {\bf D66}, 094014 (2002) [hep-ph/0208141]. 

\bibitem{gtmd1}
S. Meissner, A. Metz and M. Schlegel, JHEP {\bf 0908} 056 (2009).
\bibitem{gtmd2}
 S. Meissner, A. Metz, M. Schlegel and K. Goeke, JHEP {\bf 0808}, 038 (2008).
\bibitem{gtmd3}
 C. Lorce, B. Pasquini and M. Vanderhaeghen, JHEP {\bf 1105},  041 (2011). 
\bibitem{gtmd4}
 X. d. Ji, Phys. Rev. Lett. {\bf 91},  062001 (2003). 
\bibitem{gtmd5} 
A. V. Belitsky, X. d. Ji and F. Yuan, Phys. Rev. {\bf D69}, 074014 (2004). 
\bibitem{gtmd6} 
A. V. Belitsky and A. V. Radyushkin, Phys. Rept. {\bf 418}, 1 (2005) (and references therein). 

\bibitem{mv} 
  L.~D.~McLerran and R.~Venugopalan,
  Phys.\ Rev.\ {\bf D49}, 2233 (1994); Phys. Rev. {\bf D49}, 3352 (1994); {\it ibid}. {\bf 50}, 2225 (1994).

\bibitem{bapp}
 K. Dusling, F. Gelis, T. Lappi and R. Venugopalan, Nucl. Phys. {\bf A836}, 159 (2010) [arXiv:0911.2720]. 

\bibitem{baryonpairs} 
J. Adam et.l. (ALICE collaboration), arXiv:1612.08975.


\bibitem{urs} 
R. Baier, A. Kovner, M. Nardi  and U. Wiedemann,  Phys. Rev. {\bf D72} (2005) 094013 [hep-ph/0506126]. 

\bibitem{cut} 
A. Kovner, M. Lublinsky and H. Weigert,  Phys. Rev. {\bf D74} (2006) 114023 [hep-ph/0608258].  

\end{thebibliography}
\end{document}